\documentclass{ws-ijmpb}
\usepackage{graphicx,rotating,subfigure,amsmath,amsfonts,amssymb,delarray}
\usepackage[super,sort]{cite}

\newcommand{\be}{\begin{equation}}
\newcommand{\ee}{\end{equation}}
\newcommand{\bea}{\begin{eqnarray}}
\newcommand{\eea}{\end{eqnarray}}

\begin{document}

\markboth{Authors' Names}
{Instructions for Typing Manuscripts (Paper's Title)}

\catchline{}{}{}{}{}
%
%%%%%%%%%%%%%%%%%%%%%%%%%%%%%%%%%%%%%%%%%%%%%%%%%%%%%%%%%%%%%%%%%%%%

\title{LONG TIME CORRELATIONS OF  NONLINEAR LUTTINGER LIQUIDS}

\author{RODRIGO G. PEREIRA}

\address{Instituto de F\'{i}sica de S\~ao Carlos, Universidade de S\~ao Paulo, C.P. 369\\
 S\~ao Carlos, SP,  13560-970, Brazil\\
rpereira@ifsc.usp.br }

\maketitle

\begin{abstract}

An overview is given of the limitations of Luttinger liquid theory in describing the real time equilibrium dynamics  of critical one-dimensional systems with nonlinear dispersion relation.  After exposing the singularities of perturbation theory in band curvature effects that break the Lorentz invariance of the Tomonaga-Luttinger model, the origin of  high frequency oscillations in the long time behaviour of correlation functions is discussed. The notion that correlations decay exponentially at finite temperature is challenged by the effects of diffusion in the density-density correlation   due to umklapp scattering in lattice models.  

\end{abstract}

\section{Introduction}

When Haldane coined the name ``Luttinger liquid'',\cite{haldane} the   point was to emphasize   the universality of the theory beyond the exactly solvable model studied by Tomonaga and Luttinger.\cite{tomonaga,luttinger}  Indeed, the mapping from interacting fermions to noninteracting bosons that renders the Tomonaga-Luttinger (TL) model  solvable depends crucially on the approximation of a linear dispersion relation for   low energy excitations. However, the  thermodynamic properties predicted by Luttinger liquid (LL) theory are asymptotically exact in the low energy limit for generic critical one-dimensional systems  because perturbations associated with band curvature are irrelevant in  the renormalization group sense. This explains why Luttinger liquid behaviour is observed in so many  different systems, such as quantum wires, carbon nanotubes, spin chain compounds and cold atoms in optical lattices.\cite{giamarchi}

Neglecting irrelevant perturbations, the elementary excitation of the one-component TL model is a free boson with   massless relativistic dispersion $\omega_q=v|q|$. As a result, the  model is Lorentz invariant: Its correlation functions are preserved by continuous rotations between space  and Euclidean time. More precisely, the TL model is invariant under local conformal transformations in $(1+1)$ dimensions. This allows one to bring in the arsenal of conformal field theory (CFT), with central charge $c=1$, to compute correlation functions.\cite{cardy} Conformal invariance dictates that for large distances $x$ and long real times $t$     the ground state correlation function for a given  field $\Phi(x)$  decays as a power law 
\begin{equation}
\langle\Phi(x,t)\Phi^\dagger(0,0) \rangle\sim \frac{1}{(x-vt)^{2\Delta_+}(x+vt)^{2\Delta_-}},\label{CFTresult}
\end{equation}
where $\Delta_+$ and $\Delta_-$ are  conformal dimensions. The conformal dimensions of physical operators are determined by the Luttinger parameter $K$ of the TL model. Both the velocity $v$ and the Luttinger parameter $K$ can be extracted from the finite size spectrum in general, or calculated exactly for integrable models in particular.\cite{izergin} The result   can be generalized to models with more than one gapless degree of freedom, such as the Hubbard model away from half-filling.\cite{frahm} Using the conformal mapping to the cylinder geometry with compactified Euclidean time direction, one can   predict that at finite temperatures correlation functions decay exponentially in both $x$ and $t$.\cite{korepin} 

Perhaps due to the remarkable success of CFT methods in LL physics, the validity of Eq. (\ref{CFTresult}) is often overstated. The truth is, for any model in the LL universality class where the dispersion relation is not exactly linear, \emph{the CFT result for time-dependent correlation functions does not give the correct long time behaviour}   for real time $t>|x|/v$, i.e. inside the light cone.
While there are examples of gapless one-dimensional systems to which CFT techniques clearly do not apply --- for instance  spinful fermions in the spin-incoherent Luttinger liquid regime\cite{zvonarev} or the ferromagnetic Bose gas\cite{bosegas} --- the   general   reason for the  breakdown of LL theory in real time dynamics is the effect of band curvature.  Although formally irrelevant, perturbations to the TL model that take the form of boson decay processes generate singular contributions  to dynamical correlation functions. The solution to this quandary gave birth to the subject of ``nonlinear Luttinger liquids''\cite{science} (see Ref. 12 for a detailed review). Some predictions of the field theory for nonlinear LLs have been recently confirmed by an exact form factor approach \cite{maillet}.

The purpose of this chapter is to provide an overview of the breakdown of LL theory in equilibrium dynamics and discuss its consequences for the long time behaviour of correlation functions in critical one-dimensional systems. We will mainly focus on two aspects: the contribution  of high energy modes to correlation functions at zero temperature and the diffusive contribution due to umklapp processes which dominates the long time tail at finite temperatures.

\section{Breakdown of Luttinger liquid theory by band curvature effects}

Fermi liquid theory breaks down in one dimension because scattering between two disconnected Fermi points always leads to singularities in the particle-hole and particle-particle channels, making quasi-particles unstable as the \emph{fermion} self-energy diverges.\cite{voit} Similarly, LL theory fails to describe dynamical response functions because the \emph{boson} self-energy due to  band curvature terms is singular. However,  the singularity in this case is   connected with the macroscopic degeneracy of states comprised of multiple bosons with the same chirality, which is an artifact of the linear dispersion approximation in the TL model.

Consider the TL model  for spinless fermions (following the standard $g$-ology notation)\cite{giamarchi}\begin{equation}
H=\sum_{r=R,L}\int dx \left[v_F:\psi^\dagger_r(-ir\partial_x) \psi^{\phantom\dagger}_r:+\frac{g_2}{2}:(\psi^\dagger_r\psi^{\phantom\dagger}_r)(\psi^\dagger_{-r}\psi^{\phantom\dagger}_{-r}):+\frac{g_4}{2}:(\psi^\dagger_r\psi^{\phantom\dagger}_r)^2:\right].\label{TLmodel}
\end{equation}
Here $r=R,L=\pm$ denotes right and left movers, defined from single-particle states with momentum around $\pm k_F$, and $::$ refers to normal ordering with respect to the free fermion ground state. The fermion field operator  reads $\Psi(x)\approx e^{ik_Fx}\psi_R(x)+e^{-ik_Fx}\psi_L(x)$. Bosonization\cite{giamarchi} maps the fermionic model to the Gaussian model\begin{equation}
H=\sum_{r=R,L}\int dx\, \frac{v}{2}\left[:(\partial_x\varphi_R)^2:+:(\partial_x\varphi_L)^2:\right],
\end{equation}
where the normal ordering is with respect to the vacuum of bosons and the chiral bosonic fields obey the commutation relation $[\varphi_r(x),\partial_{x^\prime}\varphi_{r^\prime}(x^\prime)]=-ir\delta_{r,r^\prime}\delta(x-x^\prime)$. For $g_2,g_4\ll v_F$, the renormalized velocity is given approximately by ${v\approx v_F-g_4/(2\pi)}+\mathcal{O}(g^2)$.  In the notation used here, the chiral fermion fields are bosonized as \begin{equation}
\psi_r(x)\approx (2\pi\alpha)^{-1/2}\exp[-i\sqrt{2\pi}(\lambda\varphi_r+\bar\lambda\varphi_{-r})],\label{mandelstam}\end{equation}
with $\lambda=(\sqrt{K}+1/\sqrt{K})/2$ and $\bar\lambda=(-\sqrt{K}+1/\sqrt{K})/2$. Here $\alpha$ is a short-distance cutoff and $K$ is the Luttinger parameter given approximately by $K\approx 1-g_2/(2\pi v_F)+\mathcal{O}(g^2)$. Within the TL model, the single fermion Green's function at zero temperature is given by $G(x,t)=-i\langle\Psi(x,t)\Psi^\dagger(0,0)\rangle=e^{ik_Fx}G_R(x,t)+e^{-ik_Fx}G_L(x,t)$, where the Green's functions for the chiral fermions read\cite{luther}\begin{eqnarray}
G_r(x,t)&=&-i\langle\psi^{\phantom\dagger}_r(x,t)\psi^\dagger_r(0,0)\rangle=(r/2\pi)(x-rvt)^{-\lambda^2}(x+rvt)^{-\bar\lambda^2}.\label{GrLL}
\end{eqnarray}
The result in Eq. (\ref{GrLL}) has the form of Eq. (1) with conformal dimensions $\Delta_\pm=\frac18\left(K+\frac1K\pm2r\right)$. Taking the Fourier transform with the proper time ordering prescription one finds that the particle addition part of the single-fermion spectral function does not have a quasi-particle peak; instead it  behaves as a power law above a threshold energy, $A(k,\omega)\sim k^{\bar\lambda^2}(\omega-v\delta k)^{-2+\lambda^2}$  for $\delta k=k-k_F\ll k_F$.\cite{meden,voit2}  Importantly, the exponent in the spectral function is directly related to the singularities of $G_r(x,t)$ in Eq. (\ref{GrLL}) along the light cone $x=\pm vt$. For $g_2,g_4\ll v_F$, the interaction-dependent exponent can also be obtained by resumming logarithmic divergences in the perturbation theory in the fermion description,\cite{solyom} but the advantage of bosonization is that the fermion interactions are treated easily by the rescaling of the bosonic fields leading to  Eq. (\ref{mandelstam}).

What if we perturb the TL model with band curvature effects, which are indeed present in any real system? Let us assume, as usual, that we are allowed to truncate the Hilbert space to low energy states around the Fermi points, but now we  add a parabolic term in the   dispersion\cite{haldane}\begin{equation}
\delta H=-\frac1{2m}\int dx\sum_{r=R,L}\, :\psi^\dagger_r\partial_x^2 \psi^{\phantom\dagger}_r:,\label{deltaHbc}
\end{equation}
where $m$ is the effective mass at the Fermi level. After bosonizing and performing the Bogoliubov transformation  that diagonalizes the TL model, the band curvature term generates  two types of operators in general\begin{eqnarray}
\delta H &=&\frac{\sqrt{2\pi}}6\int dx\,\left\{\eta_-[:(\partial_x\varphi_L)^3:-:(\partial_x\varphi_R)^3:]\right.\nonumber\\
& &\left.+\eta_+[:(\partial_x\varphi_L)^2\partial_x\varphi_R:-:(\partial_x\varphi_R)^2\partial_x\varphi_L):]\right\},\label{etas}
\end{eqnarray}
where $\eta_\pm$ are coupling constants of order $1/m$ which can be calculated to lowest order by bosonization or fixed by phenomenological relations.\cite{pereira06,JSTAT} The important point is that these cubic terms spoil the solvability of the TL model since they introduce interactions between the bosonic modes. Nevertheless, we might hope they can be treated perturbatively.   To illustrate the problem with perturbation theory, it suffices to consider  the $\eta_-$ term, which does not mix the two chiral components of the bosonic field. Calculating the first-order correction to the single-fermion Green's function, one finds\cite{karimi}\begin{equation}
\frac{\delta G_r(x,t)}{G_r(x,t)}=-\frac{i\lambda^3\eta_-}{v}\left[\frac{1}{x-rvt}-\frac{x+rvt}{(x-rvt)^2}\right]-\frac{i\bar\lambda^3\eta_-}{v}\left[\frac{1}{x+rvt}-\frac{x-rvt}{(x+rvt)^2}\right].\label{deltaG}
\end{equation}
For $x\to\infty$ or $t\to\infty$, the expression in Eq. (\ref{deltaG}) decays faster   than the result for the TL model.\footnote{Actually, the correction due to $\eta_-$ vanishes  for  $t=0$. Equivalently, the correction for equal-time correlation functions calculated around Eq. (5.4) of Ref. 1 vanishes if the ``angle'' of the Bogoliubov transformation that mixes right and left movers is set to zero.}  This is expected since   simple power counting tells us that the $\eta_-$ perturbation has scaling dimension three and is irrelevant at the LL fixed point.  Hence  the argument for the universality of LL theory\cite{haldane}: If we are   interested in low-energy thermodynamic properties,  band curvature effects are harmless because they only give subleading corrections to correlation functions at large distances  $x\gg \alpha,vt$.

The problem shows up when we are interested in \emph{dynamical} response functions which depend on \emph{both} momentum and frequency, such as the spectral function $A(k,\omega)$. Remarkably, the correction in Eq. (\ref{deltaG}) is  more singular at the light cone $x=\pm vt$ than the unperturbed LL result in Eq. (\ref{GrLL}).  For the spectral function, this means that the corresponding correction $\delta A(k,\omega)/A(k,\omega)\sim \eta_-\delta k^2/(\omega-v\delta k)$ has a singular frequency dependence and diverges more strongly  at the lower threshold $\omega=v\delta k$ than the LL result. The singularity  actually gets worse at higher orders of perturbation theory in $\eta_-$, as one obtains more powers of $1/(x\pm vt)$ in $\delta G_r(x,t)$. 

%Therefore we are faced we are faced with a breakdown of Luttinger liquid theory at finite but small frequencies  $\omega \approx vk$.

\begin{figure}
\begin{center}
\includegraphics*[width=110mm]{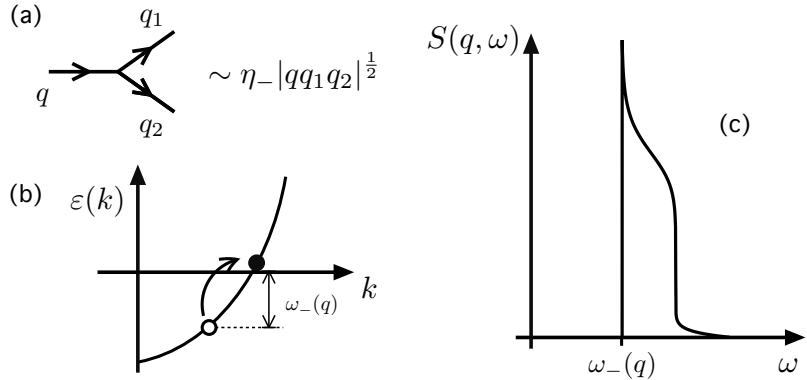}
\caption{(a) In the boson description, band curvature is represented by an interaction vertex that scales with the momenta of the bosons. But due to   Lorentz invariance of the unperturbed TL model, the  two-boson state after the decay is degenerate with the initial  single-boson state, regardless  of the value of $q_1$ (or $q_2=q-q_1$). (b) The divergences  of perturbation theory in band curvature operators can be resummed by refermionizing to new fermions with renormalized nonlinear dispersion. Density excitations are now represented by non degenerate particle-hole pairs. {(c)} Broadened boson peak in  the density structure factor $S(q,\omega)$ of a nonlinear Luttinger liquid for $q\ll k_F$. \label{fig:bandcurvature}}
\end{center}
\end{figure}

The same problem is   present in the calculation of    the density-density correlation function $\chi(x,t)=\langle n(x,t)n(0,0)\rangle$. In the boson description, the fluctuation of the fermion density operator is represented by \begin{equation}
n(x)=\,:\Psi^\dagger(x)\Psi(x):\,\approx \sqrt{K/\pi}\partial_x\phi(x)-(1/2\pi\alpha)\cos[\sqrt{4\pi K}\phi(x)-2k_Fx],\label{density}\end{equation} 
where $\phi(x)=[\varphi_L(x)-\varphi_R(x)]/\sqrt2$.  The correlation function for the long-wavelength part of $n(x)$ is equivalent to the  boson propagator\begin{equation}
\chi(x,t)\sim \sum_{r,r^\prime}\langle:\psi^\dagger_r\psi^{\phantom\dagger}_r:(x,t):\psi^\dagger_{r^\prime}\psi^{\phantom\dagger}_{r^\prime}:(0,0)\rangle=\frac{K}{\pi}\langle\partial_x\phi(x,t) \partial_x\phi(0,0)\rangle.
\end{equation}
The Fourier transform of $\chi(x,t)$ for small momentum $q\ll k_F$ yields the dynamical structure factor  $S(q,\omega)=-2\textrm{Im}\chi_{ret}(q,\omega)\approx K|q|\delta(\omega-v|q|)$. The delta function peak obtained as an approximation for $S(q\ll k_F,\omega)$ within the TL model corresponds to the spectral function of the coherent bosonic mode with well defined energy and momentum. Following the analogy with Landau's Fermi liquid theory, one would expect that going beyond the TL model and introducing band curvature effects would lead to a finite boson lifetime. However, the attempt to calculate a boson self-energy using perturbation theory fails.\cite{samokhin} Similar to   the fermion Green's function, perturbation theory in $\eta_-$ for $\chi(x,t)$ generates corrections which are increasingly more singular at the light cone $x=\pm vt$. To understand this singularity, we note that  in momentum and frequency domain  the three-legged vertex $\eta_-$  in Eq. (\ref{etas})   allows the single boson with momentum $q$ to decay into two bosons with momenta $q_1$ and $q_2=q-q_1$ (see Fig.~\ref{fig:bandcurvature}(a)); however, the energy of the intermediate state when the boson lines are put on shell is always $\omega=vq_1+vq_2=vq$, independent of the internal momenta. This huge degeneracy is present at any finite order of perturbation theory.  The result is that to any finite order the boson decay rate due to the $\eta_-$ perturbation diverges on shell  as $\sim \delta(\omega-vq)$.

The degeneracy of the many-boson states stems from the linear dispersion approximation of the TL model. In the model of free fermions with nonlinear dispersion,   particle-hole pairs with the same total momentum --- a linear combination of which defines  the bosonic excitations --- are not degenerate. Thus the cause of the breakdown of LL theory is   the impossibility of starting from the TL model  and breaking Lorentz invariance within finite order perturbation theory. The   picture of a ``quasi-boson" with well behaved self-energy due to band curvature simply does not work. It is frustrating that the  bosonization method, so helpful in resumming the divergences of Fermi liquid theory in one dimension, cannot handle the innocent looking perturbation in Eq. (\ref{deltaHbc}), which is quadratic in fermions. 

The question then is whether there is an alternative representation which captures the essential physics and resums the divergences of 
   band curvature effects in LL theory. Fortunately, the answer is yes. The trick is  to refermionize the Hamiltonian with the cubic perturbation $\eta_-$ to obtain a model of noninteracting fermions with nonlinear dispersion\cite{rozhkov,science} \begin{equation}
H+\delta H\approx  \sum_{r=R,L}\int dx :\tilde\psi^\dagger_r \left[v(-ir\partial_x)+\frac{\eta_-}{2}(-i\partial_x)^2  \right]\tilde\psi^{\phantom\dagger}_r:\label{newpsi}
\end{equation}   
The new fermions are defined such that $ :\tilde\psi^\dagger_r \tilde\psi^{\phantom\dagger}_r:=-r\partial_x\varphi_r/\sqrt{2\pi}$ and differ from the original fermions by string operators.\cite{science}  We note that the fermion interactions $g_2,g_4$ in Eq. (\ref{TLmodel}) are absorbed into the renormalization of the velocity $v$ and effective mass $\eta_-$. Including the $\eta_-$ operator as the parabolic term in the dispersion of the new fermions lifts the degeneracy of the many-boson intermediate states. The approximation in Eq. (\ref{newpsi}) is to neglect the $\eta_+$ operator defined in Eq. (\ref{etas})  as well as more irrelevant operators (with dimension four and higher).  Within this approximation,  the free-boson peak in $S(q,\omega)$ broadens into  a two-fermion continuum with rectangular line shape and width $\delta\omega_q\sim \eta_-q^2$.\cite{pustilnik,pereira06} This result is exactly what one expects from summing the infinite series of diagrams in the  $\eta_-$ perturbation.\cite{JSTAT}

The   renormalized fermion band with nonlinear dispersion relation offers a convenient starting point to study dynamical response functions. With Lorentz invariance broken at the outset,  the thresholds of the exact spectrum are not the same as in LL theory.
While  the $\eta_-$ band curvature term sets the width of the two-fermion continuum for $q\ll k_F$, additional irrelevant operators   generate interactions between the new fermions and give rise to power law singularities at the edges of the spectrum.
 Rather than governed by light cone effects, the new singularities of nonlinear Luttinger liquids are in analogy to the x-ray edge problem of optical absorption in metals.\cite{pustilnik} The lower threshold $\omega_-(q)$ below which $S(q,\omega)$ vanishes is given by the minimum energy of a particle-hole excitation with momentum $q$ created in the renormalized fermion band. For positive band curvature $\eta_->0$ as in Fig.~1(b), this corresponds to a particle at the Fermi point and a hole as deep as possible with energy $\omega_-(q) = vq-\eta_-q^2/2$.  Like in the x-ray edge problem,  this ``deep hole'' can be described as an effective quantum impurity  which propagates with different velocity than the low energy modes. The power law  singularity at the edge $\omega_-(q)$ has an exponent proportional to $q$ in the case of short range interactions.\cite{pustilnik}  Moreover, interactions between right and left movers make the spectral weight extend above the upper threshold of the two-fermion continuum $\omega_+(q)=vq+\eta_-q^2/2$ predicted by the approximation in Eq. (\ref{newpsi}) and $S(q,\omega)$  acquires a   tail that decays as $\sim \eta_+^2q^4/\omega^2$ at high frequencies $\omega-vq\gg \eta_-q^2$.\cite{pustilnik,pereira06}  
 
The picture that  emerges for the broadening of the   peak in $S(q\ll k_F,\omega)$ due to irrelevant operators is illustrated in Fig. 1{(c)}. Unlike the Lorentzian quasiparticle peak in the spectral function of Fermi liquids, the ``quasi-boson'' peak   of nonlinear Luttinger liquids is asymmetric and has an x-ray edge type singularity above the lower threshold $\omega_-(q)$. 
 
Accounting for band curvature effects, the singularities of the single fermion spectral function of Luttinger liquids are also modified.\cite{khodas}  In LL theory   particle and hole Green's functions coincide  because   the TL model is particle-hole symmetric, but in the presence of band curvature  this is no longer the case. For positive band curvature, the excitation that creates a single deep hole defines the lower threshold of the support of the hole spectral function for a given momentum $k$. As a result of kinematics, the power law singularity at the deep hole threshold cannot be broadened by any interactions. In contrast, the support of the particle spectral function does extend below the energy of the single particle excitation. In this case, three-body scattering  processes in generic (i.e. nonintegrable) models allow the single particle to decay into the continuum and a Lorentzian peak  with decay rate $\gamma_k\propto (k-k_F)^8$ is obtained.\cite{khodas}   The only surviving  power law in the particle spectral function has a positive exponent and is found at the absolute lower threshold of the support, located at the energy of the deep hole excitation.

Going back to real space and time, we should expect   the long time decay of correlation functions of Luttinger liquids to be strongly affected by the x-ray edge type singularities at the thresholds of the nonlinear spectrum. This will be the subject of the next section.

\section{High energy contributions to time-dependent correlation functions}

Once Lorentz invariance is broken,  the exponents for the long time decay of correlation functions  are not constrained to be the same as the ones for large distance decay. In fact, contrary to conventional wisdom, the long time behaviour at zero temperature is not even dominated by low energy modes, but by high energy saddle point contributions which take advantage of the dispersion nonlinearity. 

To see how this comes about, consider the simple case of free fermions with   parabolic dispersion relation $\varepsilon(k)=k^2/2m$. The hole Green's function is given exactly by\begin{equation}
G_h(x,t)=\langle\Psi^\dagger(x,t)\Psi(0,0)\rangle=\int_{-k_F}^{k_F}\frac{dk}{2\pi}\,e^{ikx-i(k^2-k_F^2)t/2m}.\label{holeGF}
\end{equation}
The real part of Eq. (\ref{holeGF}) is plotted in Fig.~2(a).  There are clearly two distinct regions in the $(x,t)$ plane. Outside the light cone, $|x|>v_Ft$, the Green's function   oscillates with distance, but not with time. This is consistent with the usual  contribution from the low energy modes which come with factors of $e^{\pm ik_Fx}$. For large distances, $|x|\gg k_F^{-1},v_Ft$, the power law decay of $G_h(x,t)$ is well described by   Eq. (\ref{GrLL}) with $K=1$. In contrast, outside the light cone, $|x|<v_Ft$, the Green's function exhibits time oscillations in addition to spatial oscillations. Such time oscillations can only come from modes with finite frequency. Indeed, a moment's reflection shows that the integral in Eq. (\ref{holeGF}) picks up significant  contributions from the saddle point away from the Fermi surface where $\frac{d}{dk}(kx-k^2t/2m)|_{k=k^*}=0$, which implies $k^*=mx/t$. In the long-time limit $v_Ft/|x|\to\infty$, the saddle point   moves to the bottom of the band, $k^*/k_F\to0$. The corresponding contribution to the hole Green's function oscillates with frequency $k_F^2/2m$ and, since the dispersion is parabolic about $k=0$,  it  decays as $1/\sqrt{t}$. The decay is slower than that of the low energy contributions, which due to the linear dispersion about $\pm k_F$ decay as $  1/t$. This dominant role of the saddle point contribution is peculiar to one dimension. In general, in $d$ dimensions, the saddle point contribution with $k\approx 0$ would decay as $t^{-d/2}$, while the Fermi surface contribution always decays as $1/t$ in the noninteracting case.\footnote{Turning on weak interactions in $d\geq 2$,   power law decay is replaced by exponential decay due to the finite lifetime of quasiparticles in Fermi liquids.}

\begin{figure}
\begin{center}
\includegraphics*[width=110mm]{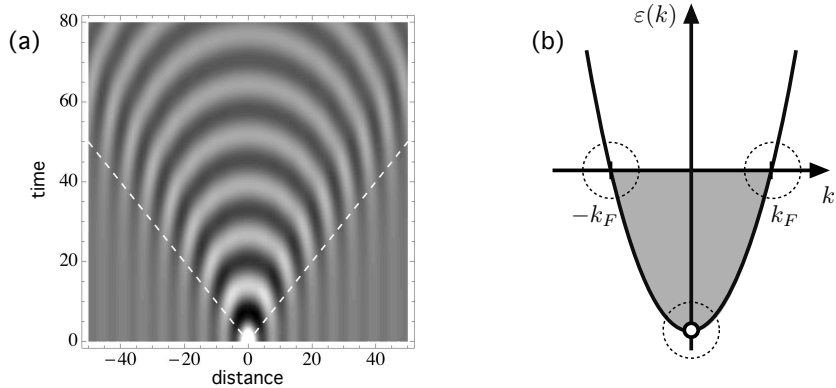}
\caption{(a) Real part of hole Green's function for free fermions with parabolic dispersion as a function of distance $x$ (in units of $k_F^{-1}$) and time $t$ (in units of $m/k_F^2$). The dashed line indicates the light cone $|x|=v_Ft$. (b) The long time behaviour inside the light cone is governed by  high energy states at the bottom of the fermion band. The effective model for the interacting case keeps states with $k\approx 0$ and $k\approx\pm k_F$.
\label{fig:timeripples}}
\end{center}
\end{figure}

In the case of parabolic dispersion there is no saddle point   above the Fermi level for $v_Ft/|x|\to\infty$, thus the long time behaviour of  the particle Green's function $G_p(x,t)=\langle\Psi(x,t)\Psi^\dagger(0,0)\rangle$ is well described by LL theory.\footnote{The problem would be the short time limit, due to the lack of a natural high energy cutoff for particle states.}  On the other hand, for   a tight-binding model  with nearest neighbour hopping $J$ the dispersion relation is $\varepsilon(k)=-2J \cos k$ (setting the lattice parameter to 1).  In this case the single-particle spectrum is bounded from above as well as from below and there is another saddle point at the top of the fermion band. Thus the particle Green's function for a lattice model also  oscillates in time  inside the light cone.

Numerical methods such as the time-dependent density matrix renormalization group (tDMRG)\cite{tdmrg} and the time-evolving block-decimation (TEBD) algorithm\cite{daley}    provide direct information about the real time evolution of one-dimensional systems. These methods   reveal that interacting models also  exhibit high frequency oscillations in equilibrium correlation functions inside the light cone  $vt>|x|$, with renormalized velocity $v$.\cite{pereira09}  Although there are no exact analytical results for     time-dependent Green's functions of interacting models, not even for integrable ones,  the exact solution for the noninteracting case suggests an approximation to describe the long time decay in  the interacting case. Besides the low energy chiral components, it is important to consider modes with  parabolic dispersion at the bottom of a renormalized fermion band. This can be achieved by pushing the  x-ray edge methods  for nonlinear Luttinger liquids\cite{pustilnik} beyond the low energy regime.\cite{pereira08} Starting from the non-interacting band, one expands the fermion field in the form $\Psi(x)\sim e^{ik_Fx}\psi_R(x)+e^{-ik_Fx}\psi_L(x)+d^\dagger(x)$, where the high energy field $d^\dagger(x)$ is defined from hole states with momentum $k\approx 0$ (see Fig.~2(b)). Using this mode expansion for the kinetic energy and interaction terms in the Hamiltonian leads to the familiar model of a mobile impurity in a  Luttinger liquid.\cite{castroneto,balents} The impurity, defined in momentum space in this case, is a single deep hole at the bottom of the band. The long time limit of the hole Green's function in the interacting case is controlled by decay of the deep hole due not only to the parabolic dispersion, but also to scattering by  low energy particle-hole pairs. The coupling between the deep hole and the low energy modes can be treated exactly within the effective impurity model using a canonical transformation that shifts the bosonic fields.\cite{schotte}  The result is that in addition to the (sub-leading) CFT terms in Eq. (\ref{GrLL}) the hole Green's function for $vt\gg|x|\gg k_F^{-1}$ has a time-oscillating term\begin{equation}
G_h(x,t)\sim \frac{e^{-iWt+iMx^2/2t}}{\sqrt{t}(v^2t^2-x^2)^{\nu/2}}.\label{highG}
\end{equation} 
Here  $W$ and $M$ are the energy and effective mass of the deep hole, respectively. The exponent for large $t$ in  $G_h(vt/|x|\to\infty)\sim e^{-iWt}/t^{\frac12+\nu}$  differs from the noninteracting result by an orthogonality catastrophe correction 
\begin{equation}\nu= \frac{1}{2K}\left(\frac\delta\pi\right)^2,
\end{equation}
where $\delta$ is interpreted as the phase shift of the Fermi surface states due to the creation of the deep hole. For a weak short-range density-density interaction $H_{int}=(1/2)\int dxdx^\prime V(x-x^\prime)n(x)n(x^\prime)$, one finds to lowest order\cite{khodas}    $\delta\approx (\tilde V_0-\tilde V_{k_F})/ v_F$, where $\tilde V_k$ is the Fourier transform of the interaction potential $V(x)$. For integrable models, it is possible to extract   phase shifts from Bethe ansatz equations and then compute the exact exponents for strong interactions.\cite{pereira08,cheianov} More generally,   phase shifts can be determined from information  about the exact high energy spectrum   using phenomenological relations.\cite{imambekov09}

The long time behaviour of   the density-density correlation function $\chi(x,t)=\langle n(x,t)n(0,0)\rangle$ also involves high energy modes. For free fermions, $\chi(x,t)$ factorizes into particle and hole Green's functions. In the case of a parabolic dispersion relation, the longest lived particle-hole excitation has total momentum $\pm k_F$ and corresponds to a hole at the bottom of band and a particle at either one of the Fermi points. With $1/\sqrt{t}$ decay for the hole Green's function and $1/t$ decay for the particle Green's function, the density-density correlation function for free fermions oscillates with frequency $k_F^2/2m$ and decays as $1/t^{3/2}$. Again, this should be compared with the decay predicted by LL theory. According to Eq. (\ref{density}), there are   low energy contributions  with momentum $q\approx 0$ (particle-hole pair around a single Fermi point) and $q\approx \pm2k_F$ (particle-hole excitation    between the two Fermi points). In the noninteracting case, both contributions decay as $1/t^2$, more rapidly than the high energy contribution. Turning on interactions  between the fermions, the parameters of the dispersion are renormalized and the exponent of  the high energy term in $\chi(x,t)$  is modified by x-ray edge type effects. The general decay is of the form\begin{equation}
\chi(x,t)\sim  \frac{e^{\pm ik_Fx-iWt+iMx^2/2t}}{\sqrt{t}(vt\mp x)^{(\lambda-\sqrt{\nu/2})^2}(vt\pm x)^{(\bar\lambda-\sqrt{\nu/2})^2}},\label{chioscillating}
\end{equation}
with $\lambda,\bar\lambda$ defined in Eq. (\ref{mandelstam}). In the long time limit, $\chi(t\gg |x|/v)\sim e^{-iWt}/t^\eta$ with exponent\cite{pereira08} \begin{equation}
\eta=\frac{1+K}{2}+\frac1{2K}\left(1-\frac{\delta}{\pi}\right)^2.
\end{equation}
In this case, the correction to the free fermion exponent  is of first order in the interaction: $\eta\approx 3/2-\delta/\pi$ for $\tilde V_0\ll v_F$. 
 The exponent $\eta$ is related to the lower edge singularity of the dynamical structure factor $S(q,\omega)$ for $q=k_F$.  The fact that $\eta$ decreases with an increasing repulsive interaction is manifested in $S(k_F,\omega)$ as a divergence at the lower edge, similar to the effect observed in the quasi-boson peak at low energies (see again Fig.~1{(c)}).

For a noninteracting  lattice model, the longest lived particle-hole excitation  is the one obtained from the saddle point contribution for both particle and hole Green's functions. This is the excitation  with total momentum $q=\pi$ that has the maximum energy allowed for a single particle-hole pair. For simplicity, let us restrict ourselves to the particle-hole symmetric case of a half-filled lattice, $k_F=\pi/2$.  The density-density correlation function picks up two factors of $e^{-iWt}/\sqrt{t}$, where $W=2J$ is half the bandwidth, from the decay of  hole and particle with parabolic dispersion.  As a result,    at large times we obtain  $\chi(t\gg |x|/v_F)\sim e^{-i2Wt}/t$. However, it turns out that this contribution is strongly suppressed by repulsive  interactions. The reason is that the problem of two high energy particles (or a particle and hole) at the threshold of a   continuum where there is an inverse square-root divergence in the joint density of states  --- related to the slow $1/t$ decay of $\chi(x,t)$ in real space and time --- is analogous to the exciton problem in one dimension. For arbitrarily weak interactions, resonant scattering between the two particles  removes the divergence at the threshold of  the density of states.\cite{mahan} Consequently,  the  exponent must change discontinuously when the fermion  interaction is switched on. For the integrable model with nearest neighbour repulsion, $\tilde V(q)=4J\Delta \cos q$ with $0<\Delta<1$ (which is equivalent to XXZ spin chain with anisotropy parameter $\Delta$), it is verified\cite{pereira08}  that the $1/t$ decay turns into a $1/t^2$ decay for times $t\gg   1/J\Delta^2$.  In  non-integrable models  the suppression must be even stronger  because this contribution is connected  with the upper threshold of the particle-hole continuum in frequency domain  and the power law at this threshold is broadened by coupling to the continuum of multiple particle-hole pairs.\cite{khodas} In real time, this implies an exponential decay for times larger than the corresponding decay rate. The conclusion is that also for    lattice models  with repulsive fermion interactions the long time behaviour of $\chi(x,t)$ is governed by the excitation with one single high energy particle (or hole) and   described by  Eq. (\ref{chioscillating}).

Summarizing this section, at zero temperature  correlation functions of critical one-dimensional systems with nonlinear dispersion oscillate at large times and decay  as power laws with non-universal exponents. The exponents depend not only on the Luttinger parameter   but also on phase shifts associated with high energy modes  with parabolic dispersion. Calculating the exact exponents requires information about the exact spectrum. There are, however, exceptions where the exponents do assume   universal values because  they are constrained by the high symmetry of the model.\cite{imambekov09} For example, in spin chains with $SU(2)$ symmetry the phase shift $\delta$ is fixed to $\delta=\pi/2$ and the oscillating term in the time-dependent spin  correlation function decays as $e^{-iWt}/t$ independently of   details of the interactions (this includes nonintegrable models with finite  range spin exchange interactions and even the Haldane-Shastry model\cite{zirnbauer}  with $1/r^2$ long-range interactions). On the other hand, at the $SU(2)$ symmetric point the Luttinger parameter becomes\cite{affleck} $K=1/2$ and the staggered ($q=2k_F=\pi$) low energy contribution decays as $1/t^{2K}=1/t$ at large times, i.e. with the same exponent as the high energy contribution. Moreover, for $SU(2)$ symmetric models one should also expect logarithmic corrections in the long time decay due to marginally irrelevant couplings between high energy modes and low energy $SU(2)$ currents.\cite{carmelo}

\section{Long time decay at finite temperatures}

Conformal invariance implies that in the TL model   the correlation function  for a field $\Phi(x)$    with conformal dimensions $(\Delta_+,\Delta_-)$ decays at finite temperatures as\cite{cardy}\begin{equation}
\langle \Phi(x,t)\Phi^\dagger(0,0)\rangle \sim\left[\frac{\pi T/v}{\sinh\pi T (x/v-t)}\right]^{2\Delta_+}\left[\frac{\pi T/v}{\sinh\pi T (x/v+t)}\label{Tcylinder}\right]^{2\Delta_-},
\end{equation}
where $T$ is the temperature. For values of $x$ and $vt$ which are small compared to the inverse temperature, $|x\pm vt|\ll v/T$, one observes the  power law decay characteristic of zero temperature correlations. For $|x\pm vt|\gg v/T$, the correlation function decays exponentially  $\sim e^{-2\pi \Delta_+T(x/v-t)}e^{-2\pi \Delta_-T(x/v+t)}$ with a thermal correlation length $\xi \sim v/T$.

As discussed in the previous section, for models with nonlinear dispersion the slowest decaying term in correlation functions involves the propagator of a  deep hole coupled to low energy modes. In Eqs. (\ref{highG}) and (\ref{chioscillating}), for example, the long time decay is  determined by the factor $e^{-iWt}/\sqrt t$ from the Green's function of the free hole with energy $W$ and parabolic dispersion   together with the factors from  low energy chiral  fields.  The  conformal dimensions of the latter are determined after the canonical transformation that decouples the  deep hole.   For $T\ll W$, the main effect of thermal fluctuations is to replace the low energy factors in the   result from the effective impurity model by the corresponding finite temperature expressions according to Eq.  (\ref{Tcylinder}). As a result, the oscillating term in time-dependent correlation functions also decays exponentially within a  time scale $\sim 1/T$. For instance, for the density-density correlation function one gets    \begin{equation}
\chi( t\gg T^{-1},x/v )\sim  \frac{e^{\pm ik_Fx-iWt}}{\sqrt{t}}e^{-\pi T(\eta-1/2)t} .\label{finiteTdecay}
\end{equation}
The exponential decay in real time is connected with the thermal broadening of x-ray edge singularities in frequency domain. Beyond the expression in Eq. (\ref{finiteTdecay}), one should also include the   decay of the high energy hole due to three-body scattering  processes, which lead to relaxation time $  \tau_h \sim W/T^2$.\cite{karzig}

However, the long time behaviour of low energy correlators can also be affected by irrelevant operators. Since irrelevant operators introduce interactions between bosonic modes, the interesting possibility is that inelastic collisions  lead to diffusive behavior in   Luttinger liquids.  The word diffusion is used here in the sense of phenomenological theories\cite{martin} for many-body systems at high temperatures  which predict that the  autocorrelation for the density of a  globally conserved quantity decay as $1/t^{d/2}$ in $d$ dimensions due to scattering-dominated random walk of the excitations. In one dimension, this means a $1/\sqrt t$ decay, which is clearly much slower than  the exponential decay predicted by scaling form in Eq. (\ref{Tcylinder}). 

\begin{figure}
\begin{center}
\includegraphics*[width=90mm]{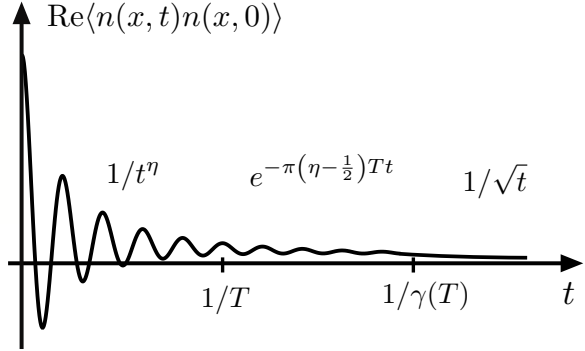}
\caption{Schematic time decay of    density autocorrelation function  $\chi(t)=\langle n(x,t)n(x,0)\rangle$ for weakly interacting lattice model at half filling   and low temperature $T$. For $t\ll 1/T$, $\chi(t)$ oscillates and decays as a power law. For $t\gtrsim 1/T$, the high frequency contribution decays exponentially. For $t\gg 1/\gamma(T)$, the low energy contribution becomes diffusive, $\chi(t)\sim T \sqrt{\gamma(T)/t}$.
\label{fig:timeripples}}
\end{center}
\end{figure}

A finite boson lifetime that gives rise to diffusion is indeed found for lattice models at half-filling.\cite{sirker,sirker2} At half-filling, particle-hole symmetry rules out cubic band curvature terms. In this case the leading perturbations to the TL model are quartic band curvature operators, which have scaling dimension four, and the non-oscillating umklapp term, which has scaling dimension $4K$. For the integrable model with nearest neighbour interactions only, the coupling constants for these perturbations are known exactly.\cite{lukyanov} While  the calculation of the boson self-energy at zero temperature is again plagued by on-shell singularities, at finite temperature the calculation  is well behaved in the regime $|\omega\pm vq|\ll T$. The result\cite{sirker} is that band curvature operators only contribute to the real part of the self-energy. The finite temperature decay rate is due entirely to   umklapp  scattering $\delta H_u=\lambda \int dx\cos(4\sqrt{\pi K}\phi)$. Computing the self-energy to  first order  in band curvature and second order in umklapp, the long wavelength part  of the density-density correlation function (i.e. the boson propagator) becomes\footnote{This expression assumes that the model is far from an integrable point and there is no ballistic channel contribution to   $\chi(q,\omega)$.\cite{sirker2} } \begin{equation}
\chi_{ret}(q,\omega)=\frac{A(T)q^2/\pi}{\omega^2-v^2(T)q^2+i2\gamma(T) \omega}.\label{damping}
\end{equation}
Here $A(T)\approx vK/[1+b(T)]$ and $v(T)\approx v[1+c(T)-b(T)]^{1/2}$ with $b(T),c(T)\sim\mathcal{O}(T^2,T^{8K-4})$  determined by the real part of the boson self-energy.  The decay  rate   is given by\cite{sirker,sirker2}\begin{equation}
\gamma(T)\approx \lambda^2\left(\frac{2\pi}{v}\right)^{8K-2} \frac{K}{2^{8K-2}} \cos^2\left(2\pi K\right)\Gamma^2\left(2 K\right)\Gamma^2\left(\frac{1}2-2K\right)T^{8K-3}.
\end{equation}
The imaginary part of Eq. (\ref{damping}) yields a Lorentzian peak with width $\gamma(T)\ll T$ for all $K>1/2$.  Particularly at the value $K=1/2$, for which the umklapp operator becomes marginal, the decay rate picks up logarithmic corrections and we obtain $\gamma(T)\sim T/\ln^2(W/T)$.  We stress that this Lorentzian approximation is only valid in the regime where $\gamma(T)$ dominates over the $T=0$ broadening due to band curvature for finite $q$. Using Eq. (\ref{damping}) to calculate the correlation function in real space and time, one finds that inside the light cone $t>|x|/v$ there is an additional contribution to $\chi(x,t)$ besides the standard CFT terms. For times $t\gg 1/\gamma(T)\gg |x|/v$, this contribution can be calculated analytically and reads\cite{sirker}\begin{equation}
\chi(x,t)\approx \frac{K T}{v^2}\sqrt{\frac{2\gamma(T)}{\pi t}}\,e^{-\gamma x^2/2v^2t}
\end{equation}
This is precisely the $1/\sqrt{t}$ decay expected for classical diffusion in one dimension, even though the assumptions of phenomenological theories definitely do not hold for Luttinger liquids. %For generic models, we expect that the diffusive term is only present in $\chi(x,t)$, which is the correlation function for the density of  a conserved quantity, namely the total number of fermions. 
As a result, the long time behaviour of the density-density correlation function  at finite temperatures  is governed by low energy modes, since the high energy terms die out exponentially for $t\gtrsim 1/T$. Notice that the important effect here is the correction to scaling due to the irrelevant umklapp operator, not to band curvature operators that break Lorentz invariance.  The real time decay of the autocorrelation function $\chi(x=0,t)$ for $T\neq0$ is illustrated in Fig.~3.

The mechanism of diffusion in the density correlator is important to explain the spin-lattice relaxation rate observed in spin-1/2 chains.\cite{thurber}  It is also connected with the question of ballistic versus diffusive transport in integrable one-dimensional systems.\cite{sirker2}

\section{Conclusion  and outlook}

We have revisited time-dependent correlation functions of critical one-dimensional systems  in light of recent theoretical advances that have  taken us beyond the paradigms of LL theory.   Although very successful in describing thermodynamic properties,  LL theory breaks down in the calculation of dynamical properties   in the presence of  band curvature. When treated as perturbations to the Lorentz-invariant  TL model,  formally irrelevant band curvature operators generate divergent corrections to the boson propagator.  However, it is possible to resum the divergences in perturbation theory by refermionizing the bosonic excitations  and considering particle-hole pairs in a band with renormalized nonlinear dispersion. This procedure yields a line shape for the ``quasi-boson'' peak in the dynamic  density-density response  which is remarkably  different than the quasi-particle peak in the spectral function of a Fermi liquid. 

Also due to band curvature effects, LL theory misses the leading terms in the asymptotic long time behaviour of correlation functions at zero temperature. In one dimension, high energy modes with parabolic dispersion located at the band edges give  contributions to time-dependent correlation functions which oscillate in time and decay as power laws with smaller exponents than the standard contributions from low energy modes. 

At finite temperatures, low energy and high energy contributions in general decay exponentially. However, for lattice models at half filling inelastic  umklapp scattering can give rise to diffusive behaviour of the low energy modes. When this happens, the density-density correlation function decays in time as a power law with universal exponent $\sim 1/\sqrt t$ for times much larger than the relaxation time.

It should be made clear that experimental probes which are only sensitive to low frequencies, such as measurements of the local density of states close to the Fermi level, $\rho(\omega)\sim \omega^{(K+K^{-1}-2)/2 }$, are not affected by the high energy contributions discussed here. However, the effects in  real time evolution  could be observed in experiments with ultra cold atoms in optical lattices. Coherent equilibrium dynamics can be investigated by preparing the system in the ground state and then creating a  local perturbation,  which is possible with  the development of techniques to address individual atoms.\cite{weitenberg} Time oscillations similar to the ones we discussed are also seen in numerical simulations of non-equilibrium dynamics of one-dimensional systems.\cite{punk} The oscillatory behaviour  is beyond the light cone effect predicted by  conformal field theory methods for quantum quenches,\cite{calabrese} but is  presumably interpreted  in terms of   effective band edges for highly excited states. Experiments suggest that questions about the  decay of correlations in equilibrium and non-equilibrium dynamics require a better understanding of the role played by the integrability of the model.\cite{kinoshita}

The prediction of a diffusive long time tail at finite temperature  is consistent with nuclear magnetic resonance experiments which probe the   dynamics of spin-1/2 chains.\cite{thurber} The decay rate $\gamma(T)$ has been confirmed numerically through the decay of the current-current correlation function,\cite{sirker} but not directly in the density-density correlation function in the low temperature regime.   It would be interesting to investigate the effects of spin diffusion in other correlation functions, such as the single-particle Green's function for spin-1/2 fermions. The  study of  finite temperature dynamics of Luttinger liquids should benefit from the recent progress in numerical methods.\cite{sirkerdiffusion,barthel}

\section*{Acknowledgements}

I am grateful to my collaborators   on this topic, in special I. Affleck, J.-S. Caux, J. Sirker, and S. R. White. This work is supported by CNPq grant 309234/2011-5.

\end{document}